# *Research on Credit Risk Early Warning Model of Commercial Banks Based on Neural Network Algorithm*

Yu Cheng[1,a], Qin Yang[2,b], Liyang Wang[3,c], Ao Xiang[4,d], Jingyu Zhang[5,e]

[1]The Fu Foundation School of Engineering and Applied Science, Columbia University, Operations Research, New York, NY, USA
[2]School of Integrated Circuit Science and Engineering (Exemplary School of Microelectronics), University of Electronic Science and Technology of China, Microelectronics Science and Engineering, Chengdu, Sichuan, China
[3]Washington University in St. Louis, Olin Business School, Finance, St. Louis, MO
[4]School of Computer Science & Engineering (School of Cybersecurity), University of Electronic Science and Technology of China, Digital Media Technology, Chengdu, Sichuan, China
[5]The Division of the Physical Sciences, Analytics, The University of Chicago, Chicago, IL, USA
[a]yucheng576@gmail.com, [b]yqin0709@gmail.com, [c]liyang.wang@wustl.edu, [d]xiangao1434964935@gmail.com, [e]simonajue@gmail.com

*Keywords:* Neural network algorithms; commercial banks; credit risk; early warning models

*Abstract:* In the realm of globalized financial markets, commercial banks are confronted with an escalating magnitude of credit risk, thereby imposing heightened requisites upon the security of bank assets and financial stability. This study harnesses advanced neural network techniques, notably the Backpropagation (BP) neural network, to pioneer a novel model for preempting credit risk in commercial banks. The discourse initially scrutinizes conventional financial risk preemptive models, such as ARMA, ARCH, and Logistic regression models, critically analyzing their real-world applications. Subsequently, the exposition elaborates on the construction process of the BP neural network model, encompassing network architecture design, activation function selection, parameter initialization, and objective function construction. Through comparative analysis, the superiority of neural network models in preempting credit risk in commercial banks is elucidated. The experimental segment selects specific bank data, validating the model's predictive accuracy and practicality. Research findings evince that this model efficaciously enhances the foresight and precision of credit risk management.

## 1. Introduction

With the rapid advancement of financial technology, the inherent limitations of traditional banking credit risk management methodologies are becoming increasingly evident. Commercial banks urgently require an intelligent system capable of swiftly responding to market dynamics and forecasting latent risks. Neural networks, particularly the Backpropagation (BP) neural network, owing to its remarkable nonlinear mapping prowess and learning capabilities, emerge as potent instruments for addressing this challenge. This discourse shall delve into the construction of an efficient credit risk alert model leveraging BP neural networks. It meticulously scrutinizes the current applications of various statistical models, delineating their inadequacies in handling complex nonlinear data. Building upon this foundation, it further elucidates the advantages of BP networks in risk prediction. Additionally, it systematically outlines the key steps involved in designing BP network models, ensuring not only a high degree of predictive accuracy but also maintaining sufficient flexibility to navigate the ever-evolving market landscape.

## 2. Introduction to Financial Risk Early Warning Models

### 2.1. ARMA model

The ARMA model, depicted in Figure 1, fully known as Autoregressive Moving Average model, stands as a prevalent forecasting instrument in the analysis of financial time series, notably adept in addressing and predicting short-term economic and financial variables. This model amalgamates the characteristics of Autoregressive (AR) and Moving Average (MA) models, adeptly capturing the dynamic features of time series. The autoregressive component relies on lagged values of the target variable, constituting a linear function of order p, reflecting the influence of past values of the variable itself. Meanwhile, the moving average component utilizes lagged values of forecast errors, comprising a linear function of order q, delineating the impact of random shocks on the current value [1].

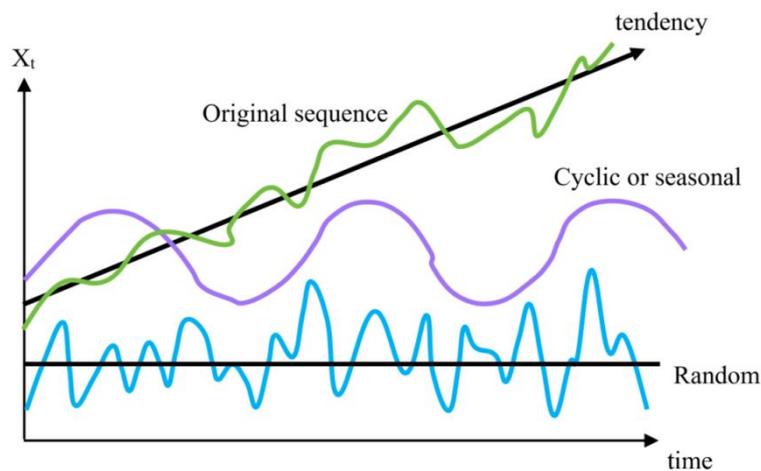

Figure 1: ARMA model

In the realm of financial risk alerting, the ARMA model finds extensive application across various economic indicators and asset prices, such as stock prices, exchange rates, and the like. Its fundamental advantage lies in its capacity to emulate the steady behavior of time series, prognosticating future trends by fitting historical data, thereby foreseeing and managing potential risks to a certain extent. Parameter estimation in the ARMA model typically employs maximum likelihood estimation or least squares estimation, necessitating the assurance of data stationarity[2]. Therefore, prior to model application, non-stationary sequences commonly undergo differencing procedures to attain stationarity before embarking on modeling analysis. Despite the ARMA model's flexibility within the financial domain, it predominantly suits dealing with relatively stable data sequences. When confronted with the high volatility and nonlinear characteristics of financial markets, the predictive performance of the ARMA model might encounter limitations. Hence, financial institutions often find it necessary to complement this model with other statistical or machine learning models to enhance the overall effectiveness of their alerting systems[3].

## 2.2. ARCH model

The ARCH model, namely the Autoregressive Conditional Heteroskedasticity model, introduced by economist Engle in 1982, is primarily utilized for modeling volatility in financial time series data. This model is particularly suited for handling financial market data with time-varying volatility, such as asset returns and exchange rates. The fundamental assumption of the ARCH model is that the variance of the current error term is a function of past error terms, distinguishing it from traditional time series models, which typically assume constant error variance. The basic expression of the ARCH model is represented as equations (1) and (2):

$$\gamma_t = bX_t + \varepsilon_t, \varepsilon_t / \Psi_{t-1} \sim N(0, \sigma_t^2) \tag{1}$$

$$\sigma_t^2 = a_0 + a_1\varepsilon_{t-1}^2 + a_1\varepsilon_{t-2}^2 + \cdots + a_p\varepsilon_{t-p}^2 \tag{2}$$

In practical application, the ARCH model adeptly captures the phenomenon of volatility clustering in financial asset returns, whereby large fluctuations are often succeeded by further large fluctuations, and small fluctuations are followed by small ones. Specifically, the variance of the error term at each moment in the model is a linear function of the square of its historical error terms. This adaptive feature allows the model to dynamically adjust risk estimates based on past volatility, thereby furnishing a dynamic early warning mechanism for risk management.

To accommodate the intricacies of financial markets, various extensions of the ARCH model have emerged, including GARCH and EGARCH, which introduce novel dynamic conditions atop the original ARCH framework to better depict and forecast the characteristics of real financial time series. For instance, the GARCH model not only incorporates the conditional variance of errors based on past error terms but also introduces a dependence on past conditional variances, thus offering a more comprehensive reflection of the impact of information sets on volatility.

## 2.3. Logistic regression modeling

The logistic regression model finds extensive application in the classification conundrums of financial risk prognosis, particularly when the target variable manifests in a binary form, such as default or non-default. This model adeptly navigates the non-linear relationships among variables and exhibits a commendable flexibility in practical deployment, unencumbered by stringent demands on data distribution shapes. Foretelling the likelihood of events, the logistic model yields outputs confined within the bounds of zero and one, wherein a predefined threshold, typically set at 0.5, delineates the transition from predictions to categorical outcomes. Ensuring the model's efficacy necessitates adherence to several fundamental prerequisites: the sampled instances must embody randomness to ensure the generalizability of analytical findings; inter-variable correlations should be tempered to avert the pernicious effects of multicollinearity, which may destabilize the model's integrity and explanatory prowess[4]; the target variable ought to manifest as a non-linear function of certain predictors, ensuring the model's aptitude for capturing intricate data structures. Despite the conspicuous advantages conferred upon logistic regression in addressing classification challenges, precise model specifications and parameter selections remain imperative for optimizing predictive performance [5]. The logistic regression model is articulated as depicted in equations (3) and (4):

$$Y = \frac{1}{1+e^{-x}} \tag{3}$$

$$\eta = C_0 + \sum_{i=1}^{m} C_i X_i \quad (i=1,2,3,...,m) \tag{4}$$

where Y ∈ [0,1], is the probability of default of the borrowing firm; Xi (0<i<p) is a financial indicator of credit risk rating
variables; Ci is the coefficients of explanatory variables, which can be obtained by regression or great likelihood estimation.

In the financial field, Logistic regression model is commonly used in credit scoring, fraud detection and other scenarios, which analyzes customers' historical data and transaction behaviors to predict their potential risky behaviors and provide scientific decision support for financial institutions. In addition, the model is more transparent and easy to interpret, which is conducive to regulatory review as well as optimization of risk management processes.

## 3. Application of BP Neural Networks in Early Warning of Operational Risks in Commercial Banks

### 3.1. Advantages of using BP neural network for commercial bank operational risk early warning

The backpropagation neural network, abbreviated as BP neural network, represents a widely employed artificial neural network within the realms of pattern recognition and data forecasting. In the context of commercial banking, the utilization of BP neural networks harbors manifold significant advantages for operational risk alertness. Initially, this network is adept at discerning and emulating intricate nonlinear relationships by assimilating copious historical data, a facet particularly pivotal in prognosticating potential operational risks within banks[6]. Subsequently, the BP neural network demonstrates commendable self-adaptation and learning prowess, continuously refining its model with the augmentation of input data, thereby enhancing the accuracy and efficacy of alerts. Furthermore, the flexible structure of BP neural networks permits the adjustment of hidden layer depths and neuron quantities in accordance with specific analytical requisites, thus enabling the model to adeptly capture nuanced features within the data, thereby facilitating more precise risk prognostication. Not only capable of handling voluminous datasets, but this network also adeptly manages noise and outliers within the data, fortifying the model's robustness [7].

In practical application, BP neural networks serve as invaluable aids for financial institutions in identifying various risk categories such as credit, market, and operational risks. Through iterative learning and optimization, they progressively elevate the scientific rigor and foresight of risk management. Consequently, the application of BP neural networks in commercial banking operational risk management not only streamlines risk control processes but also augments banks' capacity to navigate future risk dynamics adeptly[8].

### 3.2. Modeling of BP artificial neural network

#### 3.2.1. Design of network topology

Designing the network topology of a BP artificial neural network model is one of the key steps in building an effective model. This design process involves determining the number of hidden layers in the network, the number of nodes in each hidden layer, and the number of nodes in the input and output layers. In most cases, research and application experience have found that a three-layer network structure (one input layer, one hidden layer, and one output layer) typically provides good performance and less computational complexity. In specific designs, a single hidden layer is chosen for its simplicity and efficiency. While multiple hidden layers can theoretically express more complex functions, they are also prone to overfitting and local minima problems in network training, which can affect the generalization ability and prediction accuracy of the model. Single hidden layer networks have been shown to be sufficient to approximate arbitrarily complex functions in most practical problems .

The structure of the BP network used in this paper for the commercial bank operational risk prediction model is shown in Fig. 2. Where, Xt+1,Xt+2, ... ,Xt+n-1 are the n inputs, $X'_{t+n}$ is the output, which is the predicted value of Xt+n, and the number of hidden layer nodes is r(1<r<n).

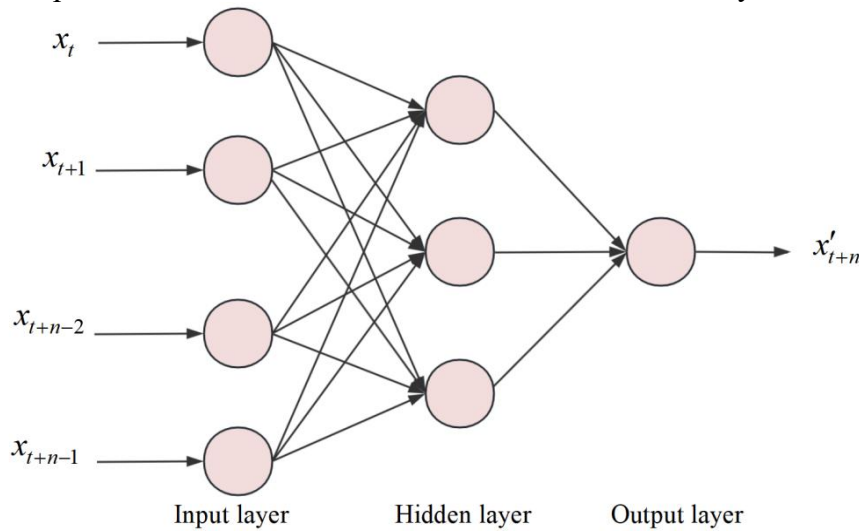

Figure 2: BP network structure of business risk prediction model for commercial banks

The number of nodes in the input and output layers is then directly determined by the number of features in the data and the output requirements of the prediction task. For example, in financial risk prediction, the number of nodes in the input layer corresponds to the number of variables used for prediction, while the output layer is usually one to several nodes depending on the number of risk types to be predicted. In addition, factors such as learning rate and activation function selection need to be considered throughout the network design process in order to construct a network model that best suits the requirements of a particular task.

### 3.2.2. Activation function selection for feedforward neural networks

When constructing artificial neural network models, the selection of appropriate activation functions stands as one of the pivotal steps. This is because activation functions wield a decisive influence over the network's learning capability and performance. The fundamental role of activation functions lies in introducing nonlinearity, thereby enabling the network to grapple with intricate data patterns. Commonly encountered activation functions are categorized into two main types: global activation functions and local activation functions.

Global activation functions, such as the Sigmoid and linear threshold units (linear functions), possess definitions and effects across the entirety of the input space. The Sigmoid function finds particular utility in scenarios where outputs necessitate confinement between 0 and 1, as observed in tasks involving probability output emulation in classification. Its advantages lie in the smoothness of output mapping, aiding in the convergence of gradient descent algorithms. Nonetheless, it also grapples with the issue of gradient vanishing, wherein gradients approach zero for either large or small input values, consequently impeding the network's learning progress.

Local activation functions, exemplified by Rectified Linear Unit (ReLU) and its variants (such as

Leaky ReLU and Parametric ReLU), only activate when inputs surpass specific threshold values. ReLU's primary merit lies in its computational simplicity and the maintenance of a constant gradient within the positive interval. This characteristic helps alleviate the problem of gradient vanishing and typically accelerates the network's learning pace. However, the ReLU function yields zero outputs for negative input values, potentially resulting in the "neuron death" issue. This phenomenon entails partial neurons outputting only zeros during the training process, thus stalling the learning of certain network structures [9]. In summary, the myriad activation functions within neural networks are depicted in Figure 3.

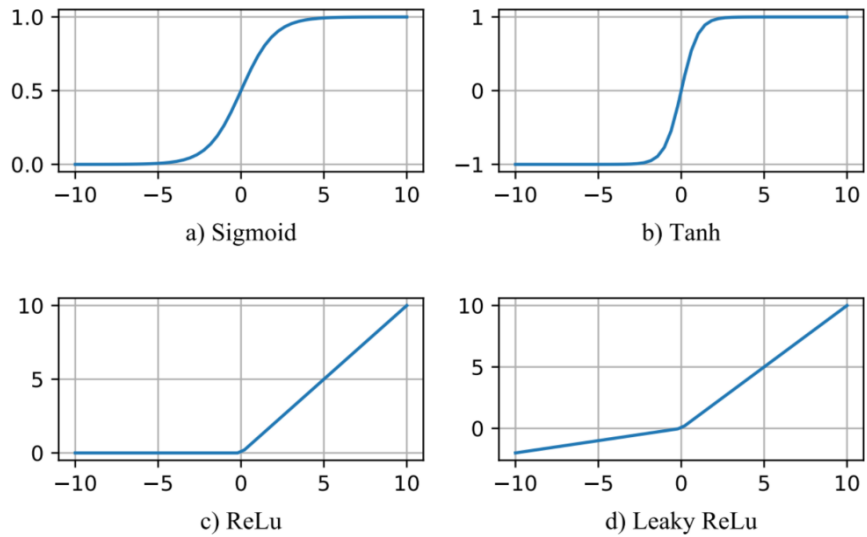

Figure 3: Activation function of neural network

## 4. Design of Credit Risk Early Warning System for Commercial Banks

In a credit risk early warning system for commercial banks, designing an effective mechanism involves a number of key components: a data processing system, a data analysis system and a decision management system, as shown in Figure 4. These three main parts work together to ensure the accuracy and efficiency of risk control[10].

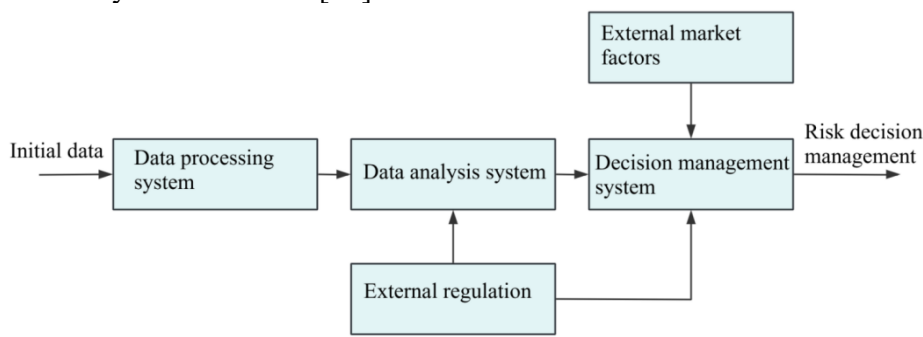

Figure 4: Operational structure of the early warning mechanism

Initially, the data processing system shoulders the responsibility of gathering, refining, and preprocessing data. It acquires personal information, transaction histories, and credit reports of loan applicants from multiple channels, purging anomalies through algorithms and formatting the data for subsequent analytical handling. Effective data processing not only enhances analytical precision but also reduces processing time, thus augmenting system responsiveness[11].

Subsequently, the data analytics system delves into the processed data through the application of statistical tests, machine learning models, and other advanced analytical techniques. This system is capable of identifying potential risk patterns and assessing borrowers' risk levels, thereby furnishing decision-making with scientific foundations. Utilizing algorithms such as decision trees, neural networks, and random forests enables accurate prediction of individual default probabilities, enhancing the precision of risk assessment.

Lastly, the decision management system constitutes the executive facet of the entire alert mechanism, formulating ultimate credit decisions based on recommendations from the data analytics system. This system not only considers customer credit scores but also comprehensively weighs market conditions, regulatory policies, and the bank's risk tolerance. Such a system ensures that banks can maintain competitiveness while controlling credit risk within acceptable bounds[12].

## 5. Empirical study on credit risk assessment model of commercial banks

### 5.1. Selection of Assessment Sample

In conducting empirical research on commercial bank credit risk assessment models, the selection of samples stands as a pivotal factor determining the quality of study and the validity of its conclusions. This study has opted for 267 distinct types of listed companies for evaluation, encompassing 34 entities under Special Treatment (ST) status and 233 non-ST companies operating under normal conditions. Such a composite sample facilitates the analysis and comparison of credit risk characteristics among companies of varying operational statuses. Through such sample selection, we can delve into the applicability and efficacy of credit risk assessment models across different types and conditions of companies, thereby necessitating essential adjustments and optimizations to the models. The diversity of samples ensures the broad applicability and high reliability of research findings, furnishing commercial banks with robust theoretical backing and data references for practical operations[13].

### 5.2. Divide the data set

In the empirical study of commercial bank credit risk assessment models, the partitioning of the dataset is crucial for both model training and validation. The study encompasses 135 samples originating from four distinct industries: electronics, chemicals, papermaking, and brewing. These samples, meticulously selected and processed, are allocated into two subsets for training and testing purposes. The training set consists of 112 samples, comprising 103 from non-ST companies and 9 from ST companies, while the testing set comprises 23 samples, with 18 from non-ST companies and 5 from ST companies. Initially, the parameters of the neural network model are set to their initial states, following which the model undergoes multiple rounds of iterative learning using the training set data. Throughout this process, the model gradually assimilates and integrates data features, optimizing network parameters to capture inherent patterns within the data, which are then stored for future use. This training methodology aims to construct a mature and reliable model capable of delving deep into the potential correlations and patterns within the dataset. Once the model training is completed, the training and testing set data are fed into this model. This not only allows for predictions of credit risk for the selected samples but also, through comparison with actual results, validates the model's accuracy and generalization ability. Such a methodological approach ensures the scientific validity and practical utility of the assessment results, providing an effective evaluation tool for bank credit risk management[14].

## 5.3. Normalization of data

In empirical studies of credit risk assessment models in commercial banks, the processing of data at the model's input layer is particularly crucial. Given the myriad evaluation indicators utilized and the vast scope of data covered, the presence of extreme values and their disparities in distribution may result in discontinuities in the model's output layer, potentially even causing derivative values to approach zero, thereby influencing the adjustment of model weight parameters. To optimize model learning speed and enhance training accuracy, the normalization of data appears especially significant.

This study employs the linear function transformation method for data normalization. Through simple mathematical transformations, this method effectively reduces the disparity between input data and set thresholds while preserving the relative relationships and characteristics of the data. Through this process, data can be compared and computed under the same standard, mitigating efficiency issues in model training caused by differences in data dimensions and value ranges. Normalized data inputted into the neural network enables smoother updates of weights and biases, reducing numerical computation errors during training and enhancing the stability and predictive accuracy of the model. This not only enhances the model's ability to handle complex data but also strengthens its adaptability and generalization to new data.

## 5.4. Network training and testing

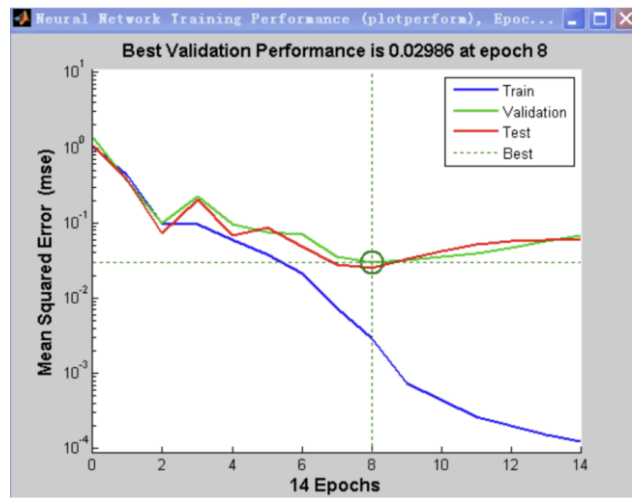

Figure 5: Neural network training results

In the present study on commercial bank credit risk assessment, the training and validation phases of the neural network model played pivotal roles. The research team initially initialized the parameters of the network model, ensuring optimal settings upon training initiation. Subsequently, leveraging the MATLAB software environment, normalized indicator data was fed into the model, facilitating thorough training of a three-layered BP neural network. Throughout the training process, the model iteratively adjusted weights and biases to optimize error rates, thereby enhancing predictive accuracy. Specific training results revealed a significant decrease in the model's error rate to below 0.01 after 8 iterations. As the number of iterations increased to 14, the error approached a marginal value of 1.0, while the model's mean squared error (MSE) reached 0.02986, indicating the model's ability to effectively capture data features and make accurate predictions. Following initial training, the model underwent further testing and validation. During the testing phase, the model's output was compared with actual data to assess its generalization capability and robustness. This process not only evaluated the model's performance on unseen data but also validated its effectiveness in practical applications, resulting in the depicted final model as illustrated in Figure 5.

The whole training and testing process reflects the meticulous steps of model tuning and the precise use of technology, which ensures the scientificity and accuracy of the credit risk assessment of commercial banks and lays a solid foundation for subsequent research and application. With this approach, the research team has successfully constructed an efficient and reliable credit risk assessment tool [15].

## 6. Conclusion

This article establishes and validates a credit risk early warning model for commercial banks based on BP neural networks, marking a significant advancement in the practice of risk management in the financial domain. Through thorough examination and comparison of traditional financial models with neural networks, this study not only enhances the accuracy of risk early warning but also furnishes banks with an efficient decision support tool, aiding them in maintaining stability amidst fierce market competition. The empirical research segment, driven by data, demonstrates the effectiveness and reliability of the model in practical applications, thereby providing a scientific basis for future commercial banks in the selection and adjustment of risk management strategies. Despite the preliminary achievements attained in current research, continual optimization and adjustment of neural network models are imperative for handling the increasingly complex and volatile financial environments. Future research endeavors may explore diverse model integration approaches and algorithmic enhancements to cater to the evolving demands of the global financial markets.

## References

*[1] Dai, S., Dai, J., Zhong, Y., Zuo, T., & Mo, Y. (2024). The Cloud-Based Design of Unmanned Constant Temperature Food Delivery Trolley in the Context of Artificial Intelligence. Journal of Computer Technology and Applied Mathematics, 1(1), 6–12.*
*[2] Li, S., Mo, Y., & Li, Z. (2022). Automated Pneumonia Detection in Chest X-Ray Images Using Deep Learning Model. Innovations in Applied Engineering and Technology, 1(1), 1–6.*
*[3] Li, Z. ., Yu, H., Xu, J., Liu, J., & Mo, Y. (2023). Stock Market Analysis and Prediction Using LSTM: A Case Study on Technology Stocks. Innovations in Applied Engineering and Technology, 2(1), 1–6.*
*[4] Yuhong Mo, Shaojie Li, Yushan Dong, Ziyi Zhu, & Zhenglin Li. (2024). Password Complexity Prediction Based on RoBERTa Algorithm. Applied Science and Engineering Journal for Advanced Research, 3(3), 1–5.*
*[5] Huang C, Bandyopadhyay A, Fan W, Miller A, Gilbertson-White S (2023) Mental toll on working women during the COVID-19 pandemic: An exploratory study using Reddit data. PLoS ONE 18(1): e0280049.*
*[6] Tongyue F ,Jiexiang X ,Zehan Z , et al.How Green Credit Policy Affects Commercial Banks' Credit Risk?: Evidence and Federated Learning-Based Modeling From 26 Listed Commercial Banks in China[J].Journal of Cases on Information Technology (JCIT),2023,26(1):18-21.*
*[7] Zhang Y, Makris Y. Hardware-based detection of spectre attacks: a machine learning approach[C]//2020 Asian hardware oriented security and trust symposium (AsianHOST). IEEE, 2020: 1-6.*
*[8] Zhang Y, Tao F, Liu X, et al. Short Video-based Advertisements Evaluation System: Self-Organizing Learning Approach[J]. arXiv preprint arXiv:2010.12662, 2020.*
*[9] Zhang Y. Hardware-based Malware Detection in Modern Microprocessors: Formal and Statistical Methods in System-level Security Assurance[D]. 2023.*
*[10] Song X, Wu D, Zhang B, et al. Zeroprompt: Streaming acoustic encoders are zero-shot masked lms[J]. arXiv preprint arXiv:2305.10649, 2023.*
*[11] Li S, Dong X, Ma D, et al. Utilizing the LightGBM Algorithm for Operator User Credit Assessment Research[J]. arXiv preprint arXiv:2403.14483, 2024.*
*[12] Li P, Abouelenien M, Mihalcea R. Deception Detection from Linguistic and Physiological Data Streams Using Bimodal Convolutional Neural Networks[J]. arXiv preprint arXiv:2311.10944, 2023.*
*[13] Li P, Lin Y, Schultz-Fellenz E. Contextual hourglass network for semantic segmentation of high resolution aerial imagery[J]. arXiv preprint arXiv:1810.12813, 2018.*
*[14] Zhang J, Xiang A, Cheng Y, et al. Research on detection of floating objects in river and lake based on AI intelligent image recognition [J]. arXiv preprint arXiv:2404.06883, 2024.*

*[15] Xiang A, Zhang J, Yang Q, et al. Research on Splicing Image Detection Algorithms Based on Natural Image Statistical Characteristics[J]. arXiv preprint arXiv:2404.16296, 2024.*